\documentclass[aps, prd, amsmath, floats, floatfix, twocolumn, nofootinbib, superscriptaddress,showpacs]{revtex4-1}
\usepackage{graphicx}
\usepackage{color}
\usepackage{latexsym}
\usepackage[colorlinks,linkcolor=blue,citecolor=blue,urlcolor=black]{hyperref}

\newcommand{\be}{\begin{equation}}
\newcommand{\ee}{\end{equation}}
\newcommand{\bea}{\begin{eqnarray}}
\newcommand{\nn}{\nonumber}
\newcommand{\eea}{\end{eqnarray}}

\newcommand{\cb}{\textcolor{black}}
\newcommand{\cbb}{\textcolor{black}}

\begin{document}

\title{Testing exterior spacetime of the neutron star via X-ray reflection spectroscopy}

\author{M. Ghasemi-Nodehi}
\email{ghasemi@nao.cas.cn}
\affiliation{National Astronomical Observatories, Chinese Academy of Sciences,  Beijing 100012, China}

\date{\today}

\begin{abstract}
The exterior geometry of a neutron star can be approximated by relativistic multipole moments of a parametrized metric using Ernst potential formalism. 
This spacetime can be tested with electromagnetic wave observation of astrophysical black holes. 
In the present paper, I simulate X-ray reflection spectra of a thin accretion disk with future X-ray missions. 
The purpose of this work is to understand whether X-ray reflection spectroscopy can distinguish the neutron star from the Kerr solution of General Relativity. 
I found that for the higher value of spin and multipole moment parameters, there are small differences in the shape of neutron star iron lines. It is hard to distinguish neutron star iron lines from Kerr ones due to the small deviations. Also, electromagnetic wave observation of slow rotating neutron stars is marginally consistent with Kerr black holes of general relativity.
\end{abstract}


\maketitle


\section{Introduction}

General relativity (GR) has successfully passed tests of the weak field gravity regime by several experiments \cite{e1,e2,e3,e4}. There is a lack of precise observational measurement in strong gravity regimes. According to GR, the final product of gravitational collapse is Kerr black holes (BHs). According to multiple moment expansion, Kerr BHs are only characterized by mass and spin and  higher multipole moments are only a function of the first two multipole moments. If independent higher multipole moments such as quadrupole moments are measured by observational data, the compact object cannot be the Kerr solution of GR. 

Neutron stars (NSs) are fascinating astrophysical objects to test strong gravity regimes and study possible deviations from the predictions of general relativity. They have high density and strong gravity. NS exterior spacetime can be accurately approximated by multipole moments expansion. Properties of NSs can be inferred by the study of higher order multipole moments. 

Hartle and Thorne~\cite{Hartle:1968si} provided the first approach for slowly rotating objects. Their approach was based on expansion up to second order in rotation. Also there have been attempts to describe spacetime around NSs by analytic solutions of stationary, axially symmetric and also vacuum spacetime, see, for example,~\cite{a1,a2,a3,a4,a5,a6}. These solutions are not constrained to a slow rotation and the numbers of parameters that construct their geometries. 
Ernst potential formalism~\cite{Ernst} provides a powerful way to generate stationary and axisymmetric spacetime solutions in GR~\cite{p1,p2,p3,p4}. This algorithm contains many parameters. 

Reference~\cite{Pap} provides an approximate solution for the spacetime around NSs. The solution is generated by the Ernst formulation of GR. This solution is parametrized by the first five multipole moments, $M, J, M_2, S_3$ and $M_4$. The author also considered the relation between these hairs and NS exterior spacetime, which depends on mass $M$, spin parameter $J/M^2$, and quadrupole moments. This type of metric can well approximate the exterior spacetime of NSs.

On the other hand, X-ray reflection spectroscopy, also known as the iron line method, is a powerful technique to unveil properties of the strong gravity regime and constrain deviations from the Kerr solution of GR. More recently, there have been studies to use this technique to test the nature of astrophysical objects and constrain deviations from GR~\cite{i1,i2,i3,i4,i5,i6,i7,i8,i9,i10,i11,i12,i13,i14,i15,i16,i17}. In this method, a geometrically thin and optically thick accretion disk emits fluorescent narrow lines by absorbing hard X-ray photon of an optically thin comptonized corona. The strongest line is the iron K$\alpha$ line, which is at 6.4 keV for neutral atoms and shifts up to 6.97 keV in the case of ionized H-like iron. This emission line is a prominent feature in X-ray reflection spectra. This emission in the inner region of the accretion disk would be broadened and asymmetric due to special and general relativistic effects of compact objects, Doppler shift, gravitational redshift and light bending.

In this paper, I simulate the iron line of a NS, in which its exterior spacetime is approximated by Ernst formalism up to five multipole moments. Then, I simulate and analyze the reflection spectrum of the NS to constrain the multipole moments. \cb{The result shows a faster rotating NS with higher value multipole moments, and slowly rotating objects cannot be distinguished from the Kerr solution of GR; thus it is hard to constrain the parameters of the NS using X-ray reflection spectroscopy.}

The structure of this paper is as follows. Details of the spacetime metric are presented in section~\ref{theory}. Section \ref{xray} is devoted to X-ray reflection spectroscopy, and the data simulation is presented in section \ref{simulation}. A summary and conclusions are given  in Section~\ref{summary}. In the following sections, I will employ natural units in which $G_{\rm N} = c = 1$ and the metric signature is $(-+++)$.


\section{The spacetime around neutron stars  \label{theory}}

Reference~\cite{Pap} provides a stationary and axially symmetric solution based on Ernst potential formalism. The solution consists of the five relativistic multipole moments, $M, J, M_2, S_3,$ and $M_4,$ respectively, the mass, angular momentum, the mass quadrupole, spin octupole and the mass hexadecapole. The line element up to these five multipole moments reads
\be \label{Papp} ds^2=-f\left(dt-\omega d\varphi\right)^2+  f^{-1}\left[ e^{2\gamma} \left( d\rho^2+dz^2 \right)+ \rho^2 d\varphi^2 \right],
\ee
where metric functions $f,\;\omega,$ and $\gamma$ are given as 

\bea   f(\rho,z) &=&1-\frac{2 M}{\sqrt{\rho ^2+z^2}}+\frac{2 M^2}{\rho ^2+z^2}\nn\\
                                   &&\!\!\!\!\!\!+\frac{\left(M_2-M^3\right) \rho ^2-2 \left(M^3+M_2\right) z^2}{\left(\rho ^2+z^2\right)^{5/2}} \nn\\ 
                                   &&\!\!\!\!\!\!+\frac{2 z^2 \left(-J^2+M^4+2 M_2 M\right)-2 M M_2 \rho ^2}{\left(\rho ^2+z^2\right)^3}\nn\\
                                    &&\!\!\!\!\!\!+\frac{A(\rho,z)}{28 \left(\rho ^2+z^2\right)^{9/2}}+\frac{B(\rho,z)}{14 \left(\rho ^2+z^2\right)^5},\\
 \omega(\rho,z) &=&-\frac{2 J \rho ^2}{\left(\rho ^2+z^2\right)^{3/2}}   -\frac{2 J M \rho ^2}{\left(\rho ^2+z^2\right)^2}  +\frac{F(\rho,z)}{\left(\rho ^2+z^2\right)^{7/2}}\nn\\
                           && \!\!\!\!\!\!  +\frac{H(\rho,z)}{2 \left(\rho ^2+z^2\right)^4}+ \frac{G(\rho,z)}{4 \left(\rho ^2+z^2\right)^{11/2}} ,\\                             
                            \gamma(\rho,z)&=&\frac{\rho ^2 \left(J^2 \left(\rho ^2-8 z^2\right)+M \left(M^3+3 M_2\right) \left(\rho ^2-4 z^2\right)\right)}{4 \left(\rho
   ^2+z^2\right)^4}\nn\\
           &&\!\!\!\!\!\!-\frac{M^2 \rho ^2}{2 \left(\rho ^2+z^2\right)^2},                     
                                    \eea
where
\bea   A(\rho,z) &=& \left[8 \rho ^2 z^2 \left(24 J^2 M+17 M^2 M_2+21 M_4\right)\right.\nn\\
      &&\!\!\!\!\!\! +\rho ^4 \left(-10 J^2 M+7 M^5+32 M_2 M^2-21 M_4\right)\nn\\
      &&\!\!\!\!\!\!\left.+8 z^4 \left(20 J^2 M-7 M^5-22 M_2 M^2-7 M_4\right)\right] , \\
   B(\rho,z)  &=&\left[\rho ^4 \left(10 J^2 M^2+10 M_2 M^3+21 M_4M+7 M_2^2\right) \right.\nn\\
          &&\!\!\!\!\!\! +4 z^4 \left(-40 J^2 M^2-14 J S_3+7 M^6+30 M_2 M^3 \right.\nn\\
          &&\!\!\!\!\!\!\left.+14 M_4 M+7 M_2^2\right)-4 \rho ^2 z^2 \left(27 J^2 M^2-21 J S_3\right.\nn\\
   &&\!\!\!\!\!\!\left.\left.+7 M^6+48 M_2 M^3+42 M_4 M+7 M_2^2\right)\right],\\
 H(\rho,z) &=&\left[4 \rho ^2 z^2 \left(J \left(M_2-2 M^3\right)-3 M S_3\right)\right.\nn\\
                                    &&\!\!\!\!\!\!\left.+\rho ^4 \left(J M_2+3 M S_3\right)\right] \\
   G(\rho,z) &=&\left[\rho ^2 \left(J^3 \left(-\left(\rho ^4+8 z^4-12 \rho ^2 z^2\right)\right)\right.\right.\nn\\
                               &&\!\!\!\!\!\!+J M \left(\left(M^3+2 M_2\right) \rho ^4-8 \left(3 M^3+2 M_2\right) z^4\right.\nn\\
                               &&\!\!\!\!\!\!\left.+4 \left(M^3+10 M_2\right) \rho ^2 z^2\right)\nn\\
                               &&\!\!\!\!\!\!\left.\left.+M^2 S_3 \left(3 \rho ^4-40 z^4+12 \rho ^2 z^2\right)\right)\right] \\
   F(\rho,z) &=&\left[\rho ^4 \left(S_3-J M^2\right)-4 \rho ^2 z^2 \left(J  M^2+S_3\right)\right] .   \eea
As general properties of this spacetime and its validity for the NS case, a horizon is located at $\rho = 0$ and problematic properties such as singularities and closed timelike curves are well inside the surface of the NS~\cite{Pap}.

In order to use this metric for the NS case, the right set of multipole moments is needed. The first multipole moments can be expressed as
\bea
M_2 &=& -\alpha j^2 M^3\,, \nn\\
S_3 &=& -\beta j^3 M^4\,, \nn\\
M_4 &=& \gamma j^4 M^5\,,
\eea
where $M$ and $j = J/M^2$ are, respectively, mass and spin parameter. The metric reduces to the Kerr solution for the case $\alpha = \beta = \gamma = 1,$ but these parameters can be larger for the NS case~\cite{k1,k2,k3,k4}. The higher moments of a NS spacetime can be expressed in terms of mass, spin and quadrupole $M_2 = Q$~\cite{k3,k4,d1}. The relation between these three parameters can be written as~\cite{k3,k4}
\bea
y_1 &=& -0.36+ 1.48\, x^{0.65} \nn \\
y_2 &=& -4.749+0.27613\, x^{1.5146} + 5.5168\, x^{0.22229}\,,
\eea
where 
\bea
x &=& \sqrt{\alpha}\,, \nn \\
y_1 &=& \sqrt[3]{\beta}\,, \nn \\
y_2 &=& \sqrt[4]{\gamma}\,.
\eea
Thus the neutron star spacetime is determined by mass, spin parameter and parameter $\alpha$. Depending on the equation state of the NS parameter, $\alpha$ ranges from $8$ for  the maximum value of the spin parameter such as $0.5$ to $\sim1.5$ for models close to the nonrotating case~\cite{Pap}.


\section{X-ray reflection spectroscopy \label{xray}}

\begin{figure}
\begin{center}
\includegraphics[type=pdf,ext=.pdf,read=.pdf,width=8.5cm]{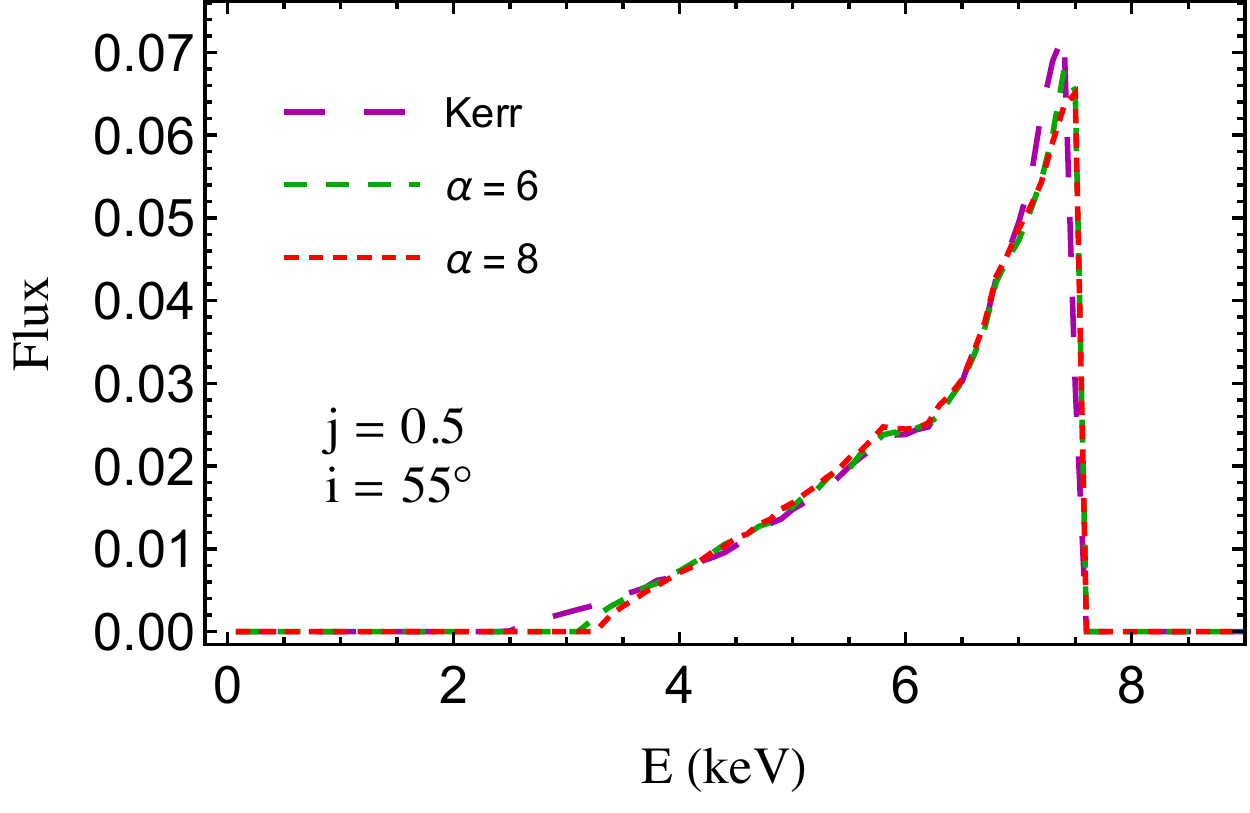}
\end{center}
\caption{The iron lines with spin parameter $0.5$ and viewing angle 55$^{\circ}$ are presented in this figure. The parameter values for $\alpha$ are $6$ and $8$. See text for more details.}
\label{fig-s05}
\end{figure}

\begin{figure}
\begin{center}
\includegraphics[type=pdf,ext=.pdf,read=.pdf,width=8.5cm]{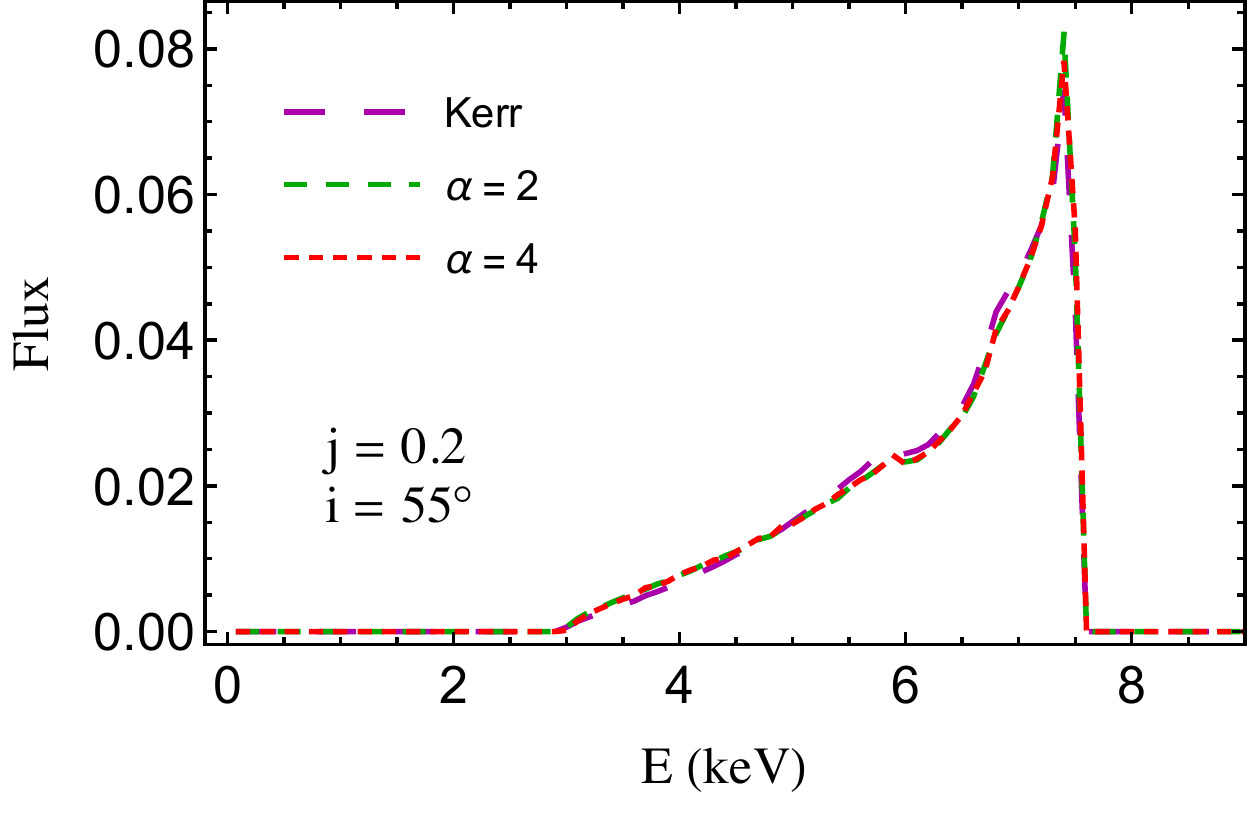}
\end{center}
\caption{Spin parameter value is $0.2$ and viewing angle is 55$^{\circ}$ in this figure. The parameter values for $\alpha$ are $2$ and $4$. See text for more details.}
\label{fig-s02}
\end{figure}

The accretion disk model in gravitational fields introduced by Novikov and Thorne is considered here~\cite{nt}. The disk is located on the equatorial plane and it is perpendicular to the black hole spin. The gas of the disk moves very nearly in a circular, geodesic orbit. The inner edge of the disk is at the innermost stable circular orbit (ISCO).  The accretion disk emits as a blackbody locally and as a multicolor blackbody when integrated radially. The corona, an optically thin electron cloud, is located at the top of the disk and may sandwich the disk. A weakly ionized iron atom in the geometrically thin and optically thick accretion disk  absorbs an X-ray photon of the hot corona and electron transition from $n = 2$ to $n = 1$ (K-shell) and releases 6.404 keV energy. This line is called the iron K$\alpha$ line. This iron K$\alpha$ line is the prominent feature in X-ray reflection spectra. This emission in the inner region of the accretion disk would be broadened and asymmetric due to the special and general relativistic effects of the compact object, Doppler shift, gravitational redshift and light bending. Studying the shape of this broad iron line can be a powerful technique to probe the strong gravity regime. This method is currently used to measure the spin of astrophysical black holes based on the Kerr metric. More recently, this method has been used to constrain possible deviations from the Kerr solution~\cite{i1,i2,i3,i4,i5,i6,i7,i8,i9,i10,i11,i12,i13,i14,i15,i16,i17}. To simulate iron lines, I use the code described in~\cite{c1,c2}. I measure the photon flux number density. The shape of the iron line depends on background metric which I consider metric~\ref{Papp} for the NS case. The iron line is also determined by the inclination angle, inner and outer radius of the disk and emissivity profile. The radius of NS is smaller than the ISCO radius. The inner edge is assumed at ISCO and the outer one is big enough not to be important. The emissivity profile is a power law, $I_e = r_e^q$, where $I_e, r_e$ and $q$ are, respectively, local intensity, radius of emission and emissivity index. Here I consider $q = -3$, which corresponds to the Newtonian limit at larger radii. I simulate iron lines for two sets with spin $0.5$ and $0.2$ based on metric~\ref{Papp} with different values of parameter $\alpha$. The inclination angle is $55^\circ$ for all cases. 
\cb{The iron lines show slight differences with the Kerr iron line for the case spin $0.5$; see Fig.~\ref{fig-s05} for simulated iron lines. }
The impact on line profile for spin value $0.2$ is weak and may be harder to constrain; see Fig.\ref{fig-s02}. In the next section, I provide data simulation and analysis to check whether these lines can be constrained or not. 


\section{Data Simulations \label{simulation}}

I simulate data to analyze reflection spectra to check possible constraints of parameters of the NS. The model is a power law and a single iron line of the code discussed in previous section. The photon index of the power law is $\Gamma = 2$, which indicates a direct component from the corona. 
\cbb{I consider the photon flux about $10^{-11}$ erg/cm$^2$/s and the iron line equivalent width to be about $200$ eV for data simulation. I first determine the normalization values of one. I then change and reset the normalizations to give the required flux and equivalent width.}
\cb{The normalization of the iron line is $3.78\times10^{-5}$ and the normalization of the power law component is 0.01. These normalizations give}
an iron line equivalent width of the simulation about $200$ eV and the photon flux about $2.6 \times 10^{-11}$ erg/cm$^2$/s in the range $2 - 10$ keV. The simulation is with a large area detector (LAD) on board Enhanced X-ray Timing and Polarimetry (eXTP) China-Europe mission, which is planned to launch before 2025 \cite{ex}. 
\cb{The LAD instrument is a set of 640 silicon drift detectors. It achieves a total effective area of about $3.4$ m$^2$ between 6 and 10 keV. It will operate in the energy range 2-30 keV and the achievable spectral resolution \cbb{is} better than 250 eV.}
The exposure time of observation is $100$ ks. The background of LAD is used to generate data with the {\it fakeit} command of XSPEC\footnote{https://heasarc.gsfc.nasa.gov/xanadu/xspec/}  software. Thus, the noises are considered. The spectra are created as 

\be
 C \left( PI \right) = \tau \int RMF\left( PI, E \right) .\,\, ARF \left( E \right) .\, S_{spec}\left( E \right) \,\, \textrm{d}E \,,
\ee

where $\tau$ is the exposure time. $RMF$ is a matrix of the response file and $ARF$ is the ancillary response file. I use $RMF$ and $ARF$ of the LAD instrument with a large effective area. $S$ is the source flux and PI is the channel. I find $C \left( PI \right)$ from the theoretical model and compare with the observed spectrum using standard fitting methods such as $\chi^2$. The reduced $\chi^2$ is
\be
\frac{\chi^2}{\nu} = \frac{1}{\nu}\sum_i^N \left( \frac{X_i - M_i}{\sigma_i} \right)^2\,,
\ee
where $X_i, M_i, \sigma_i$ and $\nu$ are, respectively, the X-ray data, expected model counts in detector channel $i$, variance and degree of freedom. I then change the parameters of the theoretical model and repeat the comparison to achieve the best fit. The LAD instrument has a large effective area to provide more counts with less Poisson noise.

\begin{figure}
\begin{center}
\includegraphics[type=pdf,ext=.pdf,read=.pdf,width=8.5cm]{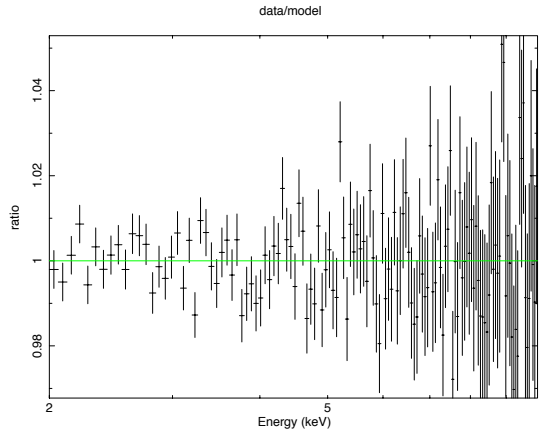}
\end{center}
\caption{\cb{Data-to-model ratio for simulation of the Kerr case with iron line $j = 0.5$, $i=55^\circ$. The reduced $\chi^2$ is $1.04$ and the fit is good. This is a test for the data simulation.}}
\label{Kerrr}
\end{figure}

\begin{figure}
\begin{center}
\includegraphics[type=pdf,ext=.pdf,read=.pdf,width=8.5cm]{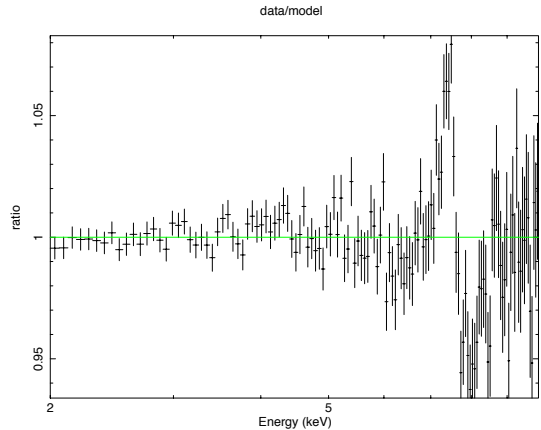}
\end{center}
\caption{Data-to-model model ratio for simulation with iron line $j = 0.5$, $i=55^\circ$ and $\alpha = 8$. The reduced $\chi^2$ is $2.1$ and there are unresolved features.}
\label{fig-s05-data}
\end{figure}

\begin{figure}
\begin{center}
\includegraphics[type=pdf,ext=.pdf,read=.pdf,width=8.5cm]{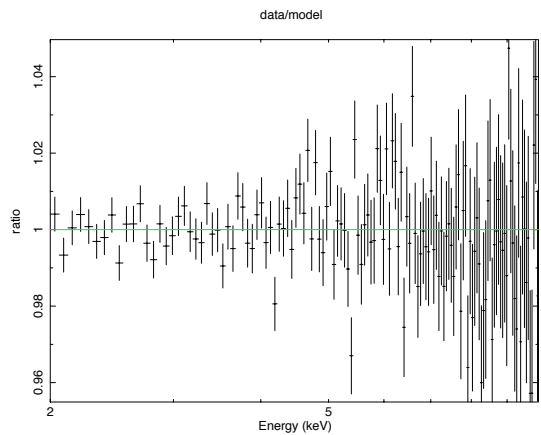}
\end{center}
\caption{The spin parameter for this figure is $0.5$, $i=55^\circ$ and $\alpha = 2$ for the iron line. The figure shows the data-to-model ratio. The reduced $\chi^2$ is $1.1$ and there are no remarkable unresolved features, which makes this case harder constrain.}
\label{fig-s02-data}
\end{figure}

I use XSPEC to fit data. The fitting model is a power law in addition to the RELLINE model. The latter model is a Kerr iron line.
\cb{Figure \ref{Kerrr} is the simulation and fitting for the Kerr case to test the simulation. The fit is good and there are no unresolved features.} 
\cb{The best fits for NS data simulation are presented in Fig.~\ref{fig-s05-data} and \ref{fig-s02-data}. Figure \ref{fig-s05-data} is for $j = 0.5$ and $\alpha = 8$. The reduced $\chi^2$ is $2.1$ and there are unresolved features. Although the fit does not seem good, as the simulated iron lines are shown slight differences with the Kerr case constraining this case might be hard due to the presence of uncertainties.}

As we saw for iron line simulations, there is not a strong impact on the iron line for the case with $j = 0.2$. The observation simulations also show that the fit is not too bad, and there are no remarkable unresolved features, which means this case is harder to constrain.

\section{Summary and conclusions \label{summary}}

Reference~\cite{Pap} provides a solution to the Einstein field equations using Ernst formalism with the first five multipole moments. This metric can be accurately related to parameters of spacetime around a NS. 

In present paper, I study iron line emission from a geometrically thin and optically thick medium around a NS. I considered two sets of spin values with different values of parameter $\alpha$. The iron line for spin value $0.5$ is \cb{slightly affected by different values of parameter $\alpha$ and also the impact of iron lines with spin value $0.2$ is not strong. }

I simulate $100$ ks observation with LAD-eXTP for the NS with different parameters values. I then fitted the simulated observations. The parameters can be constrained if the observation cannot be fitted by a Kerr model, and the fit is bad. If the fit is good, the deviation from Kerr spacetime cannot be constrained.

I found for a higher value of spin parameter, $j = 0.5$ and $\alpha =8$, \cb{the fit does not seem good, but due to small effect on simulated iron lines and presence of uncertainties, it might be hard to constrain the NS case and deviations from Kerr geometry. The fit is not too bad for  $j = 0.2$ and $\alpha =2$. This also means it is not possible to constrain possible deviations from Kerr in this case.}


\begin{acknowledgments}
I thank the Yukawa Institute for Theoretical Physics, Kyoto University, Kyoto, Japan, where this work was initiated during my visit there. I thank the School of Astronomy at the Institute for Research in Fundamental Sciences (IPM), Tehran, Iran, where part of this work was done. I am grateful to Professor Misao Sasaki for useful discussion. I also thank Professor Cosimo Bambi for reading this manuscript and providing useful comments. I acknowledge support from the China Postdoctoral Science Foundation, grant~No.~2017LH021.
\end{acknowledgments}


\end{document}